\documentclass[sigconf]{acmart}

\usepackage{times}  % DO NOT CHANGE THIS
\usepackage{helvet}  % DO NOT CHANGE THIS
\usepackage{courier}  % DO NOT CHANGE THIS
\usepackage{natbib}  % DO NOT CHANGE THIS AND DO NOT ADD ANY OPTIONS TO IT
\usepackage{caption} % DO NOT CHANGE THIS AND DO NOT ADD ANY OPTIONS TO IT
\usepackage{newfloat}
\usepackage{listings}

\usepackage[ruled, linesnumbered]{algorithm2e} % 
\usepackage{subcaption}

\usepackage{booktabs}       % professional-quality tables
\usepackage{amsfonts}       % blackboard math symbols
\usepackage{nicefrac}       % compact symbols for 1/2, etc.
\usepackage{microtype}      % microtypography
\usepackage{xcolor}         % colors
\usepackage{multirow}
\usepackage{adjustbox}

\usepackage{amsmath} % for split

\DeclareMathOperator*{\argmax}{arg\,max}

\setcopyright{acmlicensed}

\begin{document}
\fancyhead{}

\title{Revisiting Item Promotion in GNN-based Collaborative Filtering: A Masked Targeted Topological Attack Perspective}

\author{\hspace*{10mm}Yongwei Wang \quad Yong Liu \quad Zhiqi Shen}

\affiliation{%
	\institution{Nanyang Technological University, Singapore}
}
\email{{yongwei.wang, stephenliu, zqshen}@ntu.edu.sg}

\renewcommand{\shortauthors}{Wang et al.}

\begin{abstract}
Graph neural networks (GNN) based collaborative filtering (CF) have attracted increasing attention in e-commerce and social media platforms. However, there still lack efforts to evaluate the robustness of such CF systems in deployment. Fundamentally different from existing attacks, this work revisits the item promotion task and reformulates it from a targeted topological attack perspective for the first time. Specifically, we first develop a targeted attack formulation to maximally increase a target item's popularity. We then leverage gradient-based optimizations to find a solution. However, we observe the gradient estimates often appear noisy due to the discrete nature of a graph, which leads to a degradation of attack ability. To resolve noisy gradient effects, we then propose a masked attack objective that can remarkably enhance the topological attack ability. Furthermore, we design a computationally efficient approach to the proposed attack, thus making it feasible to evaluate large-large CF systems. Experiments on two real-world datasets show the effectiveness of our attack in analyzing the robustness of GNN-based CF more practically.
\end{abstract}

\begin{CCSXML}
<ccs2012>
<concept>
<concept_id>10010147.10010257.10010258.10010261.10010276</concept_id>
<concept_desc>Computing methodologies~Adversarial learning</concept_desc>
<concept_significance>500</concept_significance>
</concept>
<concept>
<concept_id>10010147.10010257.10010293.10010294</concept_id>
<concept_desc>Computing methodologies~Neural networks</concept_desc>
<concept_significance>500</concept_significance>
</concept>
<concept>
<concept_id>10010147.10010257.10010282.10011305</concept_id>
<concept_desc>Computing methodologies~Semi-supervised learning settings</concept_desc>
<concept_significance>300</concept_significance>
</concept>
<concept>
</ccs2012>
\end{CCSXML}

% \copyrightyear{2022}
% \acmYear{2022}
% \setcopyright{acmlicensed}
% \acmConference[KDD '22]{The 24th ACM SIGKDD International Conference on Knowledge Discovery \& Data Mining}{August 19--23, 2018}{London, United Kingdom}
% \acmBooktitle{KDD '22: The 24th ACM SIGKDD International Conference on Knowledge Discovery \& Data Mining, August 19--23, 2018, London, United Kingdom}
\acmPrice{15.00}
\acmDOI{10.1145/3219819.3220078}
\acmISBN{978-1-4503-5552-0/18/08}

\keywords{Item promotion attack, collaborative filtering, recommendation system, targeted topology attack, node masking }

\maketitle

%%%%%%%%% BODY TEXT
\section{Introduction}

\label{intro}
% RS; 
Collaborative filtering-based recommendation systems (RS) aim to recommend a personalized list of top-$K$ products (\textit{a.k.a} items) to each user that match best with her/his interests \cite{ekstrand2011collaborative, deldjoo2021survey}. Due to the effectiveness in promoting items, RS have been widely adopted in popular platforms, ranging from short-video discovery (\textit{e.g.,} TikTok) to e-shopping (\textit{e.g.,} Amazon). The mainstream paradigm of RS is collaborative filtering (CF), which assumes that users with similar behaviors are likely to show interests to similar items \cite{he2017neural}. As a result, CF attempts to exploit the observed user-item interactions, modeled as a user-item matrix (\textit{a.k.a} a bipartite graph), to make predictions for the unobserved ones. To better capture such interactions, graph neural networks (GNN) \cite{kipf2016semi} have attracted increasing attention in RS, and GNN-based RS achieve state-of-the-art performances in recommendation \cite{wang2019neural, he2020lightgcn}. Therefore, this work focuses on the GNN-based RS.

% considered task
Instead of trying to improve a recommender's prediction accuracy, this work investigates how to maximally boost the ranking of a low-popularity item on a potentially deployed recommendation system \cite{liu2021adversarial}. An item attains a higher popularity than another one if it is displayed in a larger number of users' recommendation list. This task finds favorable applications in scenarios where a seller expects to maximize the profits by promoting her/his items to as many potential users as possible. An intuitive solution is to encourage a number of users to show preference (\textit{i.e.,} adding positive rating) to the target item, \textit{e.g.,} by sending vouchers. However, there exist two crucial challenges to make the solution valid. The first challenge is how to ensure the newly-added ratings do contribute to the target item's popularity; and the other one is how to minimize the seller's budget (\textit{e.g.,} vouchers) by limiting the number of ratings to be added.            

\begin{figure*}[!ht]
	\centering
	\includegraphics[width=0.85\textwidth]{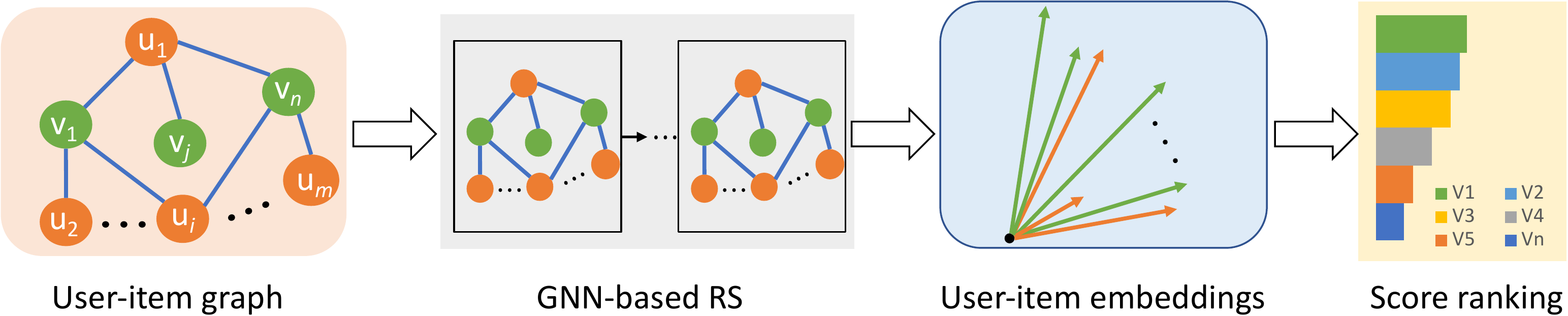}
	\caption{\textbf{Illustration of an advanced GNN-based collaborative filtering model in recommender system.} Users and items consist of a bipartite graph which is then input to a GNN-based collaborative filtering RS to generate user and item embeddings. Items that match best with a user in the embedding space will appear in the user's recommendation list. }
	\label{fig:gcn_RS}
\end{figure*}

% data poisoning
The item promotion scenario is closely related to the robustness of a collaborative filtering recommendation system. Existing works attempt to address the challenges above by creating and injecting numerous fake users into the data, a technique known as shilling attacks or data poisoning  attacks \cite{li2016data, tang2020revisiting, fang2020influence, wu2021ready}. However, these existing methods were generally specially designed for matrix factorization-based collaborative filtering recommenders, a type of conventional RS.  Thus they are inapplicable to evaluating the robustness of an advanced GNN-based collaborative filtering RS. To our best knowledge, only limited studies propose data poisoning methods that may apply for GNN-based RS \cite{tang2020revisiting, wu2021ready}. 

Unfortunately, these recently proposed methods still demand adding a large number of fake users/ratings. Besides, due to the statistical differences in rating between real and fake users, fake users may be detected and removed to mitigate the attack ability. These issues hinders attacks to take place in real scenes. Therefore, it is of urgency to develop practical and effective attacks to evaluate GNN-based collaborative filtering models in real scenes.       

% novelty part
For the first time, this work proposes a simple yet effective item promotion method on GNN-based RS from a masked topological attack perspective. The developed objective function allows us to maximize a target item's popularity with only a small number of interaction changes in the user-item graph topology. Yet it is challenging to solve the optimization problem mainly due to its combinatorial nature. To effectively address this issue, we employ a gradient-based solution and then propose a node masking mechanism to significantly enhance the attack ability. Moreover, we present a resource-efficient approach to enable our method to evaluate the robustness of large-scale GNN-based collaborative filtering systems.

Our major contributions can be summarized as follows:
\begin{itemize}
    \item This work revisits the item promotion task in a GNN-based collaborative filtering system, and re-formulates it as a targeted topological attack problem for the first time. This new formulation is fundamentally different from existing mainstream item promotion attacks in which we do not create and inject fake user profiles into the system. 
    
    \item We develop a novel node masking mechanism that can remarkably boost the attack ability of vanilla gradient-based optimization solutions. To address the memory consumption issue in large-scale graphs in real deployment, we further propose a resource-efficient approach that significantly reduces the memory cost from the quadratic to a linear one regarding the number of nodes. 
    
    \item We conduct experiments on real-world recommendation datasets with advanced GNN-based collaborative filtering models. Our results reveal that our proposed methods can substantially promote an item's popularity even given limited perturbation budgets, and it is demonstrated to consistently outperform baseline attack methods.

\end{itemize}

\section{Related Work} 
% This section will present preliminaries in GNN-based collaborative filtering, then briefly survey item promotion methods in such systems. 

\subsection{GNN-based Collaborative Filtering}
Collaborative filtering is a mainstream research in recommendation systems to predict users' preferences given collaborative signals. The essence of collaborative filtering is to learn user and item representations (\textit{a.k.a} embeddings) jointly by leveraging the user-item interaction graph \cite{wu2020graph}. Then, items will be recommended to a user whose embeddings match the best with the user's embedding. Early explorations in collaborative filtering mainly focus on the matrix-factorization (MF) model \cite{hu2008collaborative, koren2009matrix} and its variants that encode the interaction history \cite{koren2008factorization, he2018nais}. However, these methods only utilize a user's one-hop neighbors to generate the embeddings. 

Inspired by the recent progress in GNN studies that exploit multi-hop neighbors in node embedding, GNN-based collaborative filtering methods have been proposed and achieved the state-of-the-art performances. Wang et al. \cite{wang2019neural} proposed neural graph collaborative filtering (NGCF), a new collaborative filtering framework based on graph neural networks to capture the higher-order connectivity of user/item nodes. More recently, He et al. proposed LightGCN \cite{he2020lightgcn} to simplify and improve NGCF. Specifically, LightGCN removed the use of feature transformation and nonlinear activation in network design, since these two components were observed to have negative effects on model training. To supplement supervised learning, self-supervised graph learning (SGL) \cite{wu2021self} explored self-supervised learning and achieved the sate-of-the-art performance in the context of collaborative filtering to assist learning node and item representations.

\subsection{Promoting Items in Collaborative Filtering}
Although user-item interactions can effectively assist collaborative filtering, some of them may be intentionally falsified to mislead the recommender system. In the collaborative filtering regime, a most common threat is the item promotion attack \cite{li2016data, tang2020revisiting, fang2020influence, wu2021fight}, in which an attacker aims to influence a specific item recommendation list of users. More concretely, the attacker may be an incentive-driven item owner and craves to increase the chance of their own items to be recommended by a victim collaborative model.

Many existing item promotion attacks can be broadly classified into two categories: model-agnostic attacks and model-specific attacks. Model-agnostic attacks (\textit{e.g.,} RandFilter attack \cite{lam2004shilling}, average attack \cite{lam2004shilling}) do not assume knowledge of victim collaborative models, therefore they can apply to both conventional collaborative filtering models and GNN-based ones. In contrast, model-specific attacks design attacking strategies only applicable for certain types of collaborative filtering models. For example, Li et al. \cite{li2016data} formulated an integrity attack objective for MF-based models \cite{cai2010singular, jain2013low}, then solved the attack problem using gradient-based optimization methods. Fang et al. \cite{fang2020influence} proposed to utilize the influence function to select and craft fake users for top-\textit{K} MF-based recommenders. Tang et al. observed that many model-specific attacks lacked exactness in gradient computation, then proposed a more precise solution to improve the attack ability. Wu et al. designed a neural network-instantiated influence module and incorporated it into a triple generative adversarial network to craft fake users to launch attacks. 
%% note that Influential function-based attack invoves 2nd-order inverse, not scalable to large graph; but our method is scalable;
%% Moreover, existing works are all for MF-based RS, not GNN-based!
%% Give a list here? refer to our AAAI 2021 paper

Unfortunately, the model-agnostic attack methods were specially designed for MF-based models, thus they are not applicable to promoting items in GNN-based collaborative filtering RS. Meanwhile, recent studies show that graph neural networks can be vulnerable to adversarial attacks --- some unnoticeable feature or edge perturbations may significantly reduce the performance of a GNN model \cite{zugner2018adversarial, dai2018adversarial, xu2019topology, geisler2021robustness}. Intuitively, adversarial attacks can be leveraged to promote items in GNN-based recommendation models. However, these existing attacks focus on GNN-based classifiers, leaving the vulnerability of GNN-based collaborative filtering largely unexplored. 

Indeed, there are three major differences between a GNN-based classification model \cite{wu2020comprehensive} and a GNN-based collaborative filtering model \cite{wu2020graph}. First, a classification decision can be made based on the logit of one node only, while a recommendation list returned by a GNN recommender involves ranking the prediction scores of all item nodes \cite{wang2019neural, he2020lightgcn}. Therefore, unlike fooling one node only in a GNN-classifier attack, attacking a GNN recommender requires to manipulating predictions of multiple nodes simultaneously. Thus, it makes the latter case special and more challenging. Second, a node classification model consists of both edges and associative semantic features, and manipulating features can effectively enhance the attack ability \cite{zugner2018adversarial}. By contrast, a GNN recommender usually only contains user-item interactions, thus increasing the attack difficulty. Third, input graph formats and their scales are also different. The input to the former model is often a small symmetric graph, while there often includes a large bipartite graph (e.g., tens of thousands of user and item nodes) in the latter one \cite{he2020lightgcn, wu2021self}. Therefore, memory-efficient attacks remains to be developed.

\section{Methodology}
% In this section, we introduce the pipeline of the proposed item promotion method based on a masked targeted attack. We first present preliminaries of GNN-based collaborative filtering models, and then elaborate the proposed attack method.

% In this section, we present preliminary and the pipeline of the proposed item promotion attack method, and then elaborate each component in details. 

\subsection{Preliminaries}
%% GNN-based RS, and our assumptions/threat models
This study focuses on the LightGCN architecture \cite{he2020lightgcn}, a state-of-the-art backbone in GNN-based collaborative filtering models \cite{wu2021self, zhou2021selfcf, zhang2022diffusion}. 

Let $\mathcal{U}=\{ u_1, \cdots, u_M \}$ and $\mathcal{I}=\{ i_1, \cdots, i_N \}$ denote the set of $M$ users and $N$ items, respectively. Let $\mathcal{O}^{+}=\{ r_{ui} | u \in \mathcal{U}, i \in \mathcal{I} \}$ denote the interaction feedback of user $u$ for item $i$. Here we consider an implicit feedback as in many real recommenders \cite{tang2020revisiting}, i.e., $r_{ui} \in \{0, 1 \}$, where $1$ indicates a positive recommendation and $0$ means an unknown entry. Denote the user-item rating binary matrix $\mathbf{R} \in \mathbb{R}^{M \times N}$ with entries as $ r_{ui} \; (u=1, \cdots, M; \; i=1,\cdots, N)$. Then, we can construct a bipartite graph $\mathcal{G} = ( \mathcal{V}, \mathcal{E} )$, where $\mathcal{V} = \mathcal{U} \; \cup \; \mathcal{I} $ and $\mathcal{E} = \mathcal{O}^{+}$ represent the vertex (or node) set and edge set, respectively. 

GNN-based collaborative filtering leverages the user-item graph $\mathcal{G}$ to learn embeddings. To be specific, it performs neighborhood aggregations iteratively on $\mathcal{G}$ to update a node's representation \cite{he2020lightgcn, wu2021self}. The propagation rule for the $l-$th ($l=1, \cdots, L$) layer can be formally defined as,

\begin{equation}
\begin{split}
    & \mathbf{z}_{u}^{(l)} = g \bigg( \sum_{j \in \mathcal{N}_u}  \widetilde{\mathbf{R}}_{u,j} \cdot \mathbf{z}_{j}^{(l-1)} \bigg)   \\
    & \mathbf{z}_{i}^{(l)} = g \bigg( \sum_{j' \in \mathcal{N}_i} \widetilde{\mathbf{R}}^T \cdot \mathbf{z}_{j'}^{(l-1)} \bigg)
\end{split}
\label{eq:prop_rule}
\end{equation}
where $\mathbf{z}_u^{(l)} \in \mathbb{R}^d$ denotes the feature vector of user $u$ at layer $l$, \ $\mathbf{z}_u^{(0)}=\mathbf{w}_u  \in \mathbb{R}^d$ denotes the trainable and randomly initialized feature vector for user $u$, \ $g(\cdot)$ is an activation function which is often set as an identity function in recent works, $\mathcal{N}_u = \{j | (u, j) \in \mathcal{E} \}$ represents items within the neighborhood of a user $u$, and $\widetilde{\mathbf{R}}_{u, j}$ denotes the $(u, j)$-th entry of a normalized user-item rating matrix $\widetilde{\mathbf{R}}$, i.e., $\widetilde{\mathbf{R}} = \mathbf{\Lambda}_L^{-1/2} \; \mathbf{R} \; \mathbf{\Lambda}_R^{-1/2} $. Here $\mathbf{\Lambda}_L \in \mathbb{R}^{M \times M}$ is a diagonal matrix with $(u, u)$-th entry as the degree of user $u$, $\mathbf{\Lambda}_R \in \mathbb{R}^{N \times N}$ denotes a diagonal matrix with $(i, i)$-th entry as the degree of item $i$. Similarly, we have notations for item $i$'s propagation rule by changing the subscript from $u$ to $i$.  

After obtaining representations of $L$ layers, the embedding of a user (or item) node can be constructed by combining the representations computed at each layer,
\begin{equation}
    \mathbf{z}_u = f_{comb} (\mathbf{z}_u^{(l)}) , \; 
    \mathbf{z}_i = f_{comb} (\mathbf{z}_i^{(l)}) , \; \forall l \in [L]
    \label{eq:combine_embedding}
\end{equation}
where $f_{comb}(\cdot)$ denotes a representation combination function that may adopt representations from the final layer only, or utilize concatenation or weighted sum of representations from different layers \cite{wang2019neural, he2020lightgcn, wu2021self}. 

%% use embedding for prediction
In a GNN-based collaborative filtering, a typical way to obtain a recommendation prediction is by matching the embedding of a user with that of an item,
\begin{equation}
    \hat{r}_{u,i} = \big< \mathbf{z}_u, \; \mathbf{z}_i \big>
    \label{eq:recom_prediction}
\end{equation}
where $<\cdot>$ denotes an inner product, $\hat{r}_{u,i}$ is a rating score estimate that indicates how likely a user $u$ would select an item $i$. The model training can be framed into a supervised learning setting or a self-supervised learning paradigm. In deployment, a pretrained collaborative filtering model first predicts rating scores for each user, then it ranks and recommends items with top-$K$ highest scores to a user.

\subsection{Targeted Topological Attacks} \label{sec:targeted_topological_attack}
In a deployed recommender, a malicious item owner intends to promote a target item $t$ to as many users as possible, a scenario called item promotion attack. Different from existing works that attempt to craft and inject fake users to graph $\mathcal{G}$, this work formulates it from a targeted topological attack perspective. We assume the attacker (\textit{i.e.,} malicious item owner) has a white-box access to $\mathcal{G}$. We also assume that the attacker has a capability to persuade a few users to give high positive ratings to the target item (\textit{e.g.,} sending vouchers). The attacking process can be formally defined as,

\begin{equation}
    \begin{split}
         \max _{\mathbf{R}^{atk }} \; & \mathcal{L}_{atk} \Big (f_{\theta}(\mathbf{R}^{atk})_t \Big) \\
        & \textrm{s.t.} \; || \mathbf{R}^{atk} - \mathbf{R}  ||_0 \leq \Delta , \\
       & \quad \;\; \mathbf{R}^{atk } \in \{0, 1\}^{M \times N} 
    \end{split}
    \label{eq:attack_obj}
\end{equation}
where $\mathcal{L}_{atk}$ denotes an objective function to improve the ranking of a specific item $t$, $f_{\theta}$ denotes a GNN-based collaborative filtering model parameterized by $\theta$, $\mathbf{R}^{atk}$ denotes a manipulated user-item rating matrix by an attacker, $||\cdot||_0$ is an $\ell_0$ norm, and $\Delta$ represents a perturbation budget for the attack, \textit{i.e.,} the maximum number of user-item interactions allowed to manipulate.

For an arbitrary user $u \in \mathcal{U}$, denote the recommendation prediction scores for each item $i \in \mathcal{I}$ as $\boldsymbol{s}_u=[\hat{r}_{u,1},\cdots, \hat{r}_{u,N}]$. The collaborative filtering system then ranks all entries in $\boldsymbol{s}_u$ and selects top-$K$ items and recommend them to user $u$, denoted as $\Omega_K^u=[i_1^u, \cdots, i_K^u]$. Often, the target item $t$ does not lie in the recommendation set $\Omega_K^u$, thus requiring to be promoted into the set with a minimal perturbation budget.

To achieve the item promotion purpose, we formulate an objective function as,
\begin{equation}
    \mathcal{L}_{atk} = \frac{1}{M} \sum_{u \in \mathcal{U}} \Big [ \lambda \textrm{log} \sigma( \hat{r}_{u,t} ) - (1 - \lambda)  \sum_{j \in \Omega_K^u, j\neq t} \textrm{log} \sigma (\hat{r}_{u,j}) \Big ] 
    \label{eq:obj_fun_org}
\end{equation}
where $\lambda$ is a hyperparameter to balance score penalizations, $\sigma(\cdot)$ denotes a sigmoid activation function $\sigma(x)=1/ \big (1+ \textrm{exp}(-x) \big)$ that converts predicted score values to the $(0,1)$ interval.  

By substituting Eq.~(\ref{eq:obj_fun_org}) into Eq.~(\ref{eq:attack_obj}), we obtain a constraint optimization problem in the white-box targeted toplogical attack setting. 
Unfortunately, this is a combinatorial problem and finding an exact solution is NP-hard in computational complexity. Alternatively, similar to white-box adversarial attacks on images, we can leverage the gradient-based optimization to approximate the solution \cite{FGSM, madry2018towards}. 

First, we relax a discrete $\boldsymbol{R}$ as a continuous multivariable. We then compute its saliency map based on the gradients of $\boldsymbol{R}^{atk}$ with respect to each variable. The saliency map measures the contributions of every pair of user-item interactions to maximize the attack objective function in Eq.~(\ref{eq:obj_fun_org}). To satisfy the perturbation budget in Eq.~(\ref{eq:attack_obj}), we select $\Delta$ users that have highest intensity in the saliency map, and do a gradient ascent to update $\boldsymbol{R}$. Specifically, the topological attack can be expressed as,  

\begin{equation}
    \boldsymbol{R}^{atk} = \mathcal{P} \bigg ( \boldsymbol{R} + \boldsymbol{M} \odot 
    \textrm{sign} \Big(
    \nabla_{\boldsymbol{R}} \mathcal{L}_{atk} \Big )  \bigg )
    \label{eq:grad_sol_org}
\end{equation}
where $\mathcal{P}$ is a projection operator that clips the perturbed $\boldsymbol{R}$ back to the $\{0, 1\}^{M \times N}$ space, $\odot$ denotes an element-wise product, $\textrm{sign} (\cdot)$ is a sign function, $\boldsymbol{M} \in \{0, 1\}^{M \times N}$ denotes a binary mask that can be computed based on the gradient saliency map,  

\begin{equation}
    \boldsymbol{M}_{u,i} = \left \{
    \begin{aligned}
    & 1, \quad if \;  \Big ([\nabla_{\boldsymbol{R}} \mathcal{L}_{atk}]_{u,i} >0 \Big ) \cap \Big ((u,i) \in \Omega_g \Big )  \\ 
    & 0, \quad otherwise
    \end{aligned}
    \right .
    \label{eq:mask_M}
\end{equation}
where $\Omega_g$ is an index set that contains the top-$\Delta$ largest values of $\nabla_{\boldsymbol{R}}\mathcal{L}_{atk}$ and it can be formally defined as,
\begin{equation}
    \argmax_{\Omega_g \subset \mathcal{G}, |\Omega_g|=\Delta} \sum_{(u,i) \in \Omega_g} \nabla_{\boldsymbol{R}_{u,i}}\mathcal{L}_{atk}
    \label{eq:omega_g}
\end{equation}

A physical interpretation to the binary mask $\boldsymbol{M}$ is how new user-item interactions should be established in order to maximally promote a target item.  

% \subsection{Attention Masking Mechanism}
\subsection{Node Masking Mechanism}
As described in the previous section, we utilize a gradient-based optimization approach to approximate the optimal solution to Eq.~(\ref{eq:attack_obj}). Given a limited perturbation budget, we select and create user-item pair candidates that achieve highest responses in the gradient saliency map. However, the attack ability can be further improved due to potential issues in this approach. First, the gradient estimates can be noisy due to the discrete nature of variables in $\boldsymbol{R}$. Also, the quantization errors due to the utilization of the sign function may hamper the effectiveness of gradient ascent.

Notice that the derivative $ \nabla_{\boldsymbol{R}_{u,i}} \mathcal{L}_{atk}$ in Eq.~(\ref{eq:grad_sol_org}) is a summation of individual derivatives computed from the log scores of $M$ user nodes w.r.t. the binary variable $\boldsymbol{R}_{u,i}$. To negate noisy effects in gradient estimates, an intuitive way is to adopt a subset of user nodes by masking out unimportant ones. We prefer preserving nodes with high predicted scores $\hat{r}_{u,t}$ than those with lower ones $\hat{r}_{u',t}$ for the target item $t$. This is because item $t$ is more likely enter into the top-$K$ recommendation list of the user $u$ than user $u'$ after a single step gradient ascent.

We design a pre-filtering step for the node masking mechanism based on predicted scores from the GNN-based collaborative filtering system. Specifically, we compose a user subset $\mathcal{U}' \subset \mathcal{U}$ that satisfies,
\begin{equation}
   \mathcal{U}' =\Big \{  u \; | \;  u \in \mathcal{U}, \; \sigma(\hat{r}_{u',t}) \geq \gamma \Big \}
    \label{eq:prefitering}
\end{equation}
where $\gamma$ denotes a masking threshold parameter. Then, a masked objective function $\mathcal{L}_{atk}^m$ can be expressed as,
\begin{equation}
        \mathcal{L}_{atk}^m = \frac{1}{|\mathcal{U}'|} \sum_{u \in \mathcal{U}'} \Big [ \lambda \textrm{log} \sigma( \hat{r}_{u,t} ) - (1 - \lambda)  \sum_{j \in \Omega_K^u, j\neq t} \textrm{log} \sigma (\hat{r}_{u,j}) \Big ] 
    \label{eq:obj_fun_masked}
\end{equation}
Clearly, Eq.~(\ref{eq:obj_fun_org}) is a special case of Eq.~(\ref{eq:obj_fun_masked}) by setting $\gamma$ to be 0. It is worth noting that our node masking mechanism is clearly different from masked strategy used in \cite{geisler2021robustness}. First of all, the research tasks are different: our work targets an item promotion task in a collaborative filtering system that involves ranking, while work \cite{geisler2021robustness} deals with a classification task. Second, we rank predicted scores and mask out user nodes with low prediction confidence below a threshold, while work \cite{geisler2021robustness} necessitates a comparison with the groundtruth labels of nodes and removes the incorrectly classified ones. Moreover, the objective functions are fundamentally different because of different tasks.

\subsection{Scaling to Large-scale Graphs}
The gradient ascent-based solution in Eq.~(\ref{eq:grad_sol_org}) requires to compute gradients w.r.t. each entry in $\boldsymbol{R}$. This approach works well on a graphics processing unit (GPU) card with limited memory when the input user-item interaction matrix $\boldsymbol{R}$ is of relatively small size. In real deployment scenes, however, with a dense gradient $\nabla_{\boldsymbol{R}}\mathcal{L}_{atk}^m$, computational issues can arise if it involves a large-scale graph that consists of thousands even millions of user and item nodes.

\textbf{Proposition 1}. \textit{In a one-layer GNN-based collaborative filtering model defined in Eq.~(\ref{eq:prop_rule}), the partial derivatives satisfy $\nabla_{\boldsymbol{R}_{u,t}}\mathcal{L}_{atk}^m > \nabla_{\boldsymbol{R}_{u,j}}\mathcal{L}_{atk}^m, (j \neq t)$ if $<\mathbf{z}_{u}^{(0)}, \mathbf{z}_{i}^{(0)}> \; \to 1$ for $u=1,\cdots,M, i=1,\cdots,N$.
}

\textbf{Remark}. The analysis above indicates that there only necessitates to compute gradients with respect the target item $t$, i.e,, $\nabla_{\boldsymbol{R}_{u,t}} (u=1,\cdots,M)$ in Eq.~(\ref{eq:grad_sol_org}). Empirically, we observe that $\nabla_{\boldsymbol{R}_{u,t}}\mathcal{L}_{atk}^m > \nabla_{\boldsymbol{R}_{u,j}}\mathcal{L}_{atk}^m, (j \neq t)$ also holds for the multi-layer well trained GNN-based collaborative filtering models. In this way, the memory consumption can be reduced from $\mathcal{S} (M \times N)$ to $\mathcal{S} (M)$, which is a significant cost reduction especially when $N$ is a large value.  

In implementation (\textit{e.g.,} using PyTorch \cite{paszke2019pytorch}), we can split matrix $\boldsymbol{R}$ into three separate tensors: $\boldsymbol{R}=\big [ \boldsymbol{R}_{:, :t-1}, \boldsymbol{R}_{:,t}, \boldsymbol{R}_{:,t+1:N} \big ]$, where only tensor $\boldsymbol{R}_{:,t}$ requires a gradient computation. Then we do a regular forward process to compute the targeted loss as in Eq.~(\ref{eq:obj_fun_masked}), and then backpropagate gradients to tensor $\boldsymbol{R}_{:,t}$.

The algorithm of the proposed method is presented in Algorithm ~\ref{alg:masked_targeted}.

\begin{algorithm}[tb]
	\footnotesize
	\SetAlgoLined
	\KwData{A pretrained $f_{\theta}$ that consists of $\mathbf{w}_u$ and $\mathbf{w}_i$, data $\boldsymbol{R} \in \mathbb{R}^{M \times N}$, target item $t$, perturbation budget $\Delta$, parameter $\lambda$, masking threshold $\gamma$. } 
 	\KwResult{A perturbed $\boldsymbol{R}^{atk}$ that satisfies  Eq.~(\ref{eq:omega_g}).}
 	
 	\tcp{Initialization and forward}
 	Initialize embeddings $\mathbf{z}_u^{(0)}=\mathbf{w}_u $, $\mathbf{z}_i^{(0)}=\mathbf{w}_i $, set $l=1$ \;
 	Rewrite $\boldsymbol{R}$: $\boldsymbol{R} \leftarrow \big [ \boldsymbol{R}_{:, :t-1}, \boldsymbol{R}_{:,t}, \boldsymbol{R}_{:,t+1:N} \big ]$\;
 	Normalize $\boldsymbol{R}$: $\widetilde{\mathbf{R}} \leftarrow \mathbf{\Lambda}_L^{-1/2} \; \mathbf{R} \; \mathbf{\Lambda}_R^{-1/2}$ \;
 	\While{$l \leq L$}{
 	Compute users embeddings at layer $l$: $\mathbf{z}_{u}^{(l)} \leftarrow g \Big ( \boldsymbol{R}_{:, :t-1} \cdot \boldsymbol{z}_i^{l-1}[:t-1,:] +  \boldsymbol{R}_{:,t} \cdot \boldsymbol{z}_i^{l-1}[t,:] + \boldsymbol{R}_{:, t:N} \cdot \boldsymbol{z}_i^{l-1}[t:N,:] \Big) $ \;
 	Compute items embeddings at layer $l$: $\mathbf{z}_{i}^{(l)} \leftarrow \Big [ g \big (\boldsymbol{R}_{:,t}^T \cdot \boldsymbol{z}_u^{l-1} \big); g \big(\boldsymbol{R}_{:,t}^T \cdot \boldsymbol{z}_u^{l-1} \big); g \big(\boldsymbol{R}_{:, t:N}^T \cdot \boldsymbol{z}_u^{l-1} \big)  \Big ] $ \;
 	$l \leftarrow l+1$\;
 	}
 	Compute final user and item embeddings using Eq.~(\ref{eq:combine_embedding})\;
 	
 	\tcp{Backward for gradient computation}
 	Compute masked targeted loss $\mathcal{L}_{atk}^m$ using Eq.~(\ref{eq:obj_fun_masked}) \;
 	Compute $\nabla_{\boldsymbol{R}_{:,t}} \mathcal{L}_{atk}^m$ using autograd, and set the rest gradients in $\nabla_{\boldsymbol{R}}$ as 0 \;
 	Find top-$\Delta$ largest values in $\nabla_{\boldsymbol{R}}$, and compute binary mask $\boldsymbol{M}$ using Eq.~(\ref{eq:mask_M})\;
 	Compute $\boldsymbol{R}^{atk}$ using gradient ascent in Eq.~(\ref{eq:grad_sol_org})\;
 	
 	\textbf{Return}: The perturbed $\boldsymbol{R}^{atk}$.
	\caption{The proposed scalable algorithm for masked targeted attacks on GNN-based collaborative filtering models.}
	\label{alg:masked_targeted}
\end{algorithm}

\section{Experiments}
\label{experiments}
In this section, we demonstrate the effectiveness of our item promotion attack method in GNN-based collaborative filtering systems. We first introduce the experimental setup, then conduct experiments on two real-world datasets for empirical validation under different settings.    

\subsection{Experimental Setup} \label{sec:exp_setup}
\textbf{Datasets:} We conduct experiments on Gowalla \cite{cho2011friendship} and Yelp2018 \cite{yelp}, two commonly-used datasets for recommendation \cite{wang2019neural, he2020lightgcn}. For both datasets, we use the pre-processed dataset with train/test split following work \cite{he2020lightgcn}.  Gowalla contains 29,858 users and 40,981 items, with an overall number of user-item interactions as 1,027,370. Yelp2018 includes 31,667 users and 38,047 items, and has 1,561,406 user-item interactions in total. 

\textbf{Models:} We evaluate our method on the LightGCN and variant models, the state-of-the-art GNN-based collaborative filtering recommenders. LightGCN is trained on Gowalla and Yelp2018 datasets, respectively, with PyTorch implementations officially released by \cite{he2020lightgcn}. We adopt default hyperparameters as shown in the official implementation. After sufficient training, LightGCN achieves good performances on both datasets. The recommendation performances on the clean datasets are reported in Appendix.

\textbf{Evaluation Protocols:} We demonstrate the attack ability to promote a target item on low popular items on Gowalla \cite{cho2011friendship} and Yelp2018 datasets. For a well trained collaborative filtering model, an item with fewer positive ratings in $\boldsymbol{R}$ will be less popular than another that has more positive feedbacks. Therefore, we use the degree of an item to quantify the popularity. To be specific, we compose three low popular target item sets based on an item's original degree. The percentile of the three item sets are: $Q_{10}, Q_{30}, Q_{50}$, respectively. For each item from the three item sets, two perturbation budgets are defined: $\Delta_s^1=deg(Q_{65}) - deg(Q_s)$ and $\Delta_s^2= 
\bar{d} - deg(Q_s)$, where $\bar{d}$ is the mean degree, $deg(q)$ denotes the degree of an item that lies in a percentile $q$, and $s \in \{10, 30, 50 \}$. To better reflect the trend of item promotion improvements, we also adopt a continually varying number of perturbation budgets. The perturbation budgets are shown in Table~\ref{tab:perturb_budgets}.

\begin{table}[]
\centering
\begin{tabular}{|c|c|c|c|c|c|c|}
\hline
Dataset  & \multicolumn{1}{l|}{$\Delta_{10}^1$} & \multicolumn{1}{l|}{$\Delta_{10}^2$} & \multicolumn{1}{l|}{$\Delta_{30}^1$} & \multicolumn{1}{l|}{$\Delta_{30}^2$} & \multicolumn{1}{l|}{$\Delta_{50}^1$} & \multicolumn{1}{l|}{$\Delta_{50}^2$} \\ \hline
Gowalla  & 7                                    & 12                                   & 5                                    & 10                                   & 3                                    & 8                                    \\ \hline
Yelp2018 & 17                                   & 25                                   & 13                                   & 21                                   & 8                                    & 16                                   \\ \hline
\end{tabular}
\caption{Perturbation budgets for topological attacks.}
\label{tab:perturb_budgets}
\end{table}

For the quantitative measure, we utilize hit number ($HN$) to evaluate the item promotion ability. For a specific target item, $HN@50$ is defined as the frequency that users have this item to be displayed in their top-$50$ recommendation list. To be more accurate, we define a pruned hit number ($PHN@50$) metric that removes the number of newly-added users from the $HN@50$ metric. For reproducibility, we report the averaged $HN@10$ and $PHN@K$ over 30 number of randomly selected target items from three low popular item sets individually. All reported results utilize fixed hyperparameters $\lambda=0.5$, $\gamma=0.95$.

\textbf{Comparison Methods:} We utilize and modify existing model-agnostic attacks on collaborative filtering models and use them as our baseline methods (e.g., random attack \cite{lam2004shilling}). Please note that we cannot compare with recent model-specific attacks (\textit{e.g.,} \cite{li2016data, fang2020influence, tang2020revisiting}), because they were developed for MF-based CF, and they do not apply for attacking GNN-based CF models. Besides, our considered settings are dramatically different from model-specific attacks in that these methods require injecting a set of fake users into training data, while our method focuses on selecting and modifying a small number of existing users. Besides RandFilter, we also design two other heuristic attacks as our baseline attacks. The compared methods are:
\begin{itemize}
    \item \textbf{RandFilter}: RandFilter was originally used for explicit ratings \cite{lam2004shilling}, and we modify it for implicit rating. We randomly select $\Delta$ users and asked them to give positive ratings to the target item $t$.
    \item \textbf{IUFilter}: From the users perspective, users that have frequent purchasing history tend to be more influential in positive rating. Therefore we choose top-$\Delta$ such users and let them to rate positively to the target items. 
    \item \textbf{RUFilter}: RUFilter selects users that have top-$\Delta$ predicted rating scores for item $t$ and put corresponding entries in $\boldsymbol{R}$ as 1 in the implicit recommendation setting.  
\end{itemize}

\subsection{Promoting Items in White-box Scenes}
An attacker is first assumed to have a white-box access to the pretrained GNN-based CF model. The attacker can use three baseline attacks and the proposed attack (in Algorithm ~\ref{alg:masked_targeted}) to conduct item promotion attacks. Three sets of low popularity items (i.e., $Q_{10}, Q_{30}, Q_{50}$) will be evaluated with different perturbation budgets. The comparison results are reported in Table~\ref{tab:promotion_attacks}.

\begin{table*}[!htbp]
\centering
\begin{adjustbox}{width=17.5cm}
\begin{tabular}{|c|l|cccccc|}
\hline
\multirow{3}{*}{Dataset}  & \multicolumn{1}{c|}{\multirow{3}{*}{Attack}} & \multicolumn{6}{c|}{Low popularity items}                                                                                                                                                                                                                                                        \\ \cline{3-8} 
                          & \multicolumn{1}{c|}{}                        & \multicolumn{2}{c|}{$Q_{10}$}                                                                           & \multicolumn{2}{c|}{$Q_{30}$}                                                                           & \multicolumn{2}{c|}{$Q_{50}$}                                                      \\ \cline{3-8} 
                          & \multicolumn{1}{c|}{}                        & \multicolumn{1}{c|}{\textit{PHN@50 ($\Delta_{10}^1$)}} & \multicolumn{1}{c|}{\textit{PHN@50 ($\Delta_{10}^2$)}} & \multicolumn{1}{c|}{\textit{PHN@50 ($\Delta_{30}^1$)}} & \multicolumn{1}{c|}{\textit{PHN@50 ($\Delta_{30}^2$)}} & \multicolumn{1}{c|}{\textit{PHN@50 ($\Delta_{50}^1$)}} & \textit{PHN@50 ($\Delta_{50}^2$)} \\ \hline \hline 
\multirow{4}{*}{Gowalla}
% \multirow{5}{*}{Gowalla}  & None                                         & \multicolumn{2}{c|}{2.6}                                                                              & \multicolumn{2}{c|}{6.4}                                                                              & \multicolumn{2}{c|}{11.5}                                                        \\ \cline{2-8} 
                          & RandFilter                                   & \multicolumn{1}{c|}{0.7}                          & \multicolumn{1}{c|}{0.5}                          & \multicolumn{1}{c|}{2.2}                          & \multicolumn{1}{c|}{0.8}                          & \multicolumn{1}{c|}{3.9}                          & 1.8                          \\ \cline{2-8} 
                          & IUFilter                                     & \multicolumn{1}{c|}{0.5}                          & \multicolumn{1}{c|}{0.3}                          & \multicolumn{1}{c|}{1.9}                          & \multicolumn{1}{c|}{0.7}                          & \multicolumn{1}{c|}{4.5}                          & 1.5                          \\ \cline{2-8} 
                          & RUFilter                                     & \multicolumn{1}{c|}{9.1}                          & \multicolumn{1}{c|}{15.6}                         & \multicolumn{1}{c|}{14.2}                         & \multicolumn{1}{c|}{25.7}                         & \multicolumn{1}{c|}{17.1}                         & 29.0                         \\ \cline{2-8} 
                          & \textbf{Proposed}                            & \multicolumn{1}{c|}{\textbf{41.4}}                & \multicolumn{1}{c|}{\textbf{117.6}}               & \multicolumn{1}{c|}{\textbf{39.1}}                & \multicolumn{1}{c|}{\textbf{106.3}}               & \multicolumn{1}{c|}{\textbf{28.9}}                & \textbf{87.7}                \\ \hline \hline
\multirow{4}{*}{Yelp2018}
% \multirow{5}{*}{Yelp2018} & None                                         & \multicolumn{2}{c|}{1.3}                                                                              & \multicolumn{2}{c|}{1.7}                                                                              & \multicolumn{2}{c|}{4.0}                                                           \\ \cline{2-8} 
                          & RandFilter                                   & \multicolumn{1}{c|}{0}                            & \multicolumn{1}{c|}{0}                            & \multicolumn{1}{c|}{0.2}                          & \multicolumn{1}{c|}{0.1}                          & \multicolumn{1}{c|}{1.2}                          & 0.4                          \\ \cline{2-8} 
                          & IUFilter                                     & \multicolumn{1}{c|}{0}                            & \multicolumn{1}{c|}{0}                            & \multicolumn{1}{c|}{0.2}                          & \multicolumn{1}{c|}{0.1}                          & \multicolumn{1}{c|}{0.8}                          & 0.2                          \\ \cline{2-8} 
                          & RUFilter                                     & \multicolumn{1}{c|}{7.4}                          & \multicolumn{1}{c|}{19.1}                         & \multicolumn{1}{c|}{4.8}                          & \multicolumn{1}{c|}{10.5}                         & \multicolumn{1}{c|}{7.6}                          & 16.5                         \\ \cline{2-8} 
                          & \textbf{Proposed}                            & \multicolumn{1}{c|}{\textbf{60.6}}                & \multicolumn{1}{c|}{\textbf{140.8}}               & \multicolumn{1}{c|}{\textbf{32.6}}                & \multicolumn{1}{c|}{\textbf{106.7}}               & \multicolumn{1}{c|}{\textbf{20.7}}                & \textbf{75.0}                \\ \hline
\end{tabular}
\end{adjustbox}
\caption{Performance comparisons of different attacks in improving a target item's popularity on Gowalla and Yelp2018 datasets. Three low popularity item sets ($Q_{10}, Q_{30}, Q_{50}$) are used for performance evaluation with perturbation budgets as $\Delta_s^1$ and $\Delta_s^2 (s=10, 30, 50)$. $PHN@50$ is averaged over 30 randomly selected target items at each item set. The best performances are marked in bold. }
\label{tab:promotion_attacks}
\end{table*}

From Table~\ref{tab:promotion_attacks}, we can observe that our proposed method achieves the highest averaged $PHN@50$ for all settings, substantially outperforming baseline methods. For example, when target items are from $Q_{10}$ and a perturbation budget as $\Delta_{10}^1$ on Gowalla, the $PHN@50$ values are 0.7, 0.5, 9.1 for RandFilter, IUFilter and RUFilter, respectively; while our method achieves a $PHN@50$ as $\boldsymbol{41.4}$, which is $\boldsymbol{4.5 \times} $ larger than the second best method. The superiority of our method is even more prominent for target items from $Q_{10}$ with the perturbation budget as $\Delta_{10}^2$, \textit{i.e.,} $\boldsymbol{7.5 \times}$ stronger than RUFilter. Although the item promotion ability tends to decrease from $Q_{10}$ to $Q_{50}$, the performance of our method is still significantly better than all baseline methods. We can arrive at a same conclusion for experiments on Yelp2018. 

\begin{figure}[!ht]
	\centering
	\includegraphics[width=0.49\textwidth]{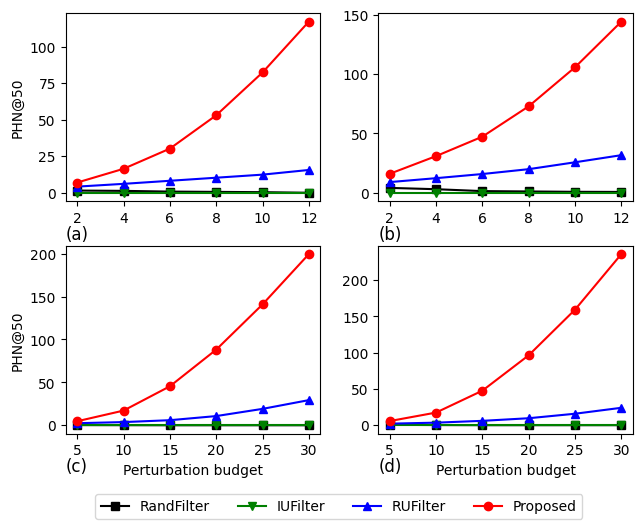}
	\caption{\textbf{Performance comparisons with a gradual varying number of budgets on low-popular items from Gowalla and Yelp2018 datasets.} (a) and (b) display $PHN@50$ results for target items from Gowalla with $Q_{10}$ and $Q_{30}$, and (c) and (d) show $PHN@50$ from Yelp2018 with $Q_{10}$ and $Q_{30}$, respectively.  }
	\label{fig:phn_budgets}
\end{figure}

In addition, we vary the perturbation budgets gradually and show in Fig.~(\ref{fig:phn_budgets}) the comparison results. Fig.~(\ref{fig:phn_budgets}) reveals that as the perturbation budgets increase, the promotion ability of our method increase dramatically, and the performance gap becomes larger compared to baseline methods. This observation indicates that GNN-based CF model is vulnerable to the proposed masked topological attack, particularly with a relatively large adversarial perturbation budget.

\subsection{Promoting Items in Black-box Scenes}
In addition to white-box attacks, we study the effectiveness of our method in a black-box setting, in which an attacker is assumed to have no knowledge to the victim models. In this setting, an attacker first adopts the pertrained model as a substitute model (sub. model) and creates a perturbed graph for a target item. The attacker then attempts to promote the target item on a unknown collaborative filtering model. Based on \cite{he2020lightgcn}, we obtain three victim models by setting different number of layers and the length of embeddings. Please refer to Appendix for detailed setup.

\begin{figure}[!ht]
	\centering
	\includegraphics[width=0.38\textwidth]{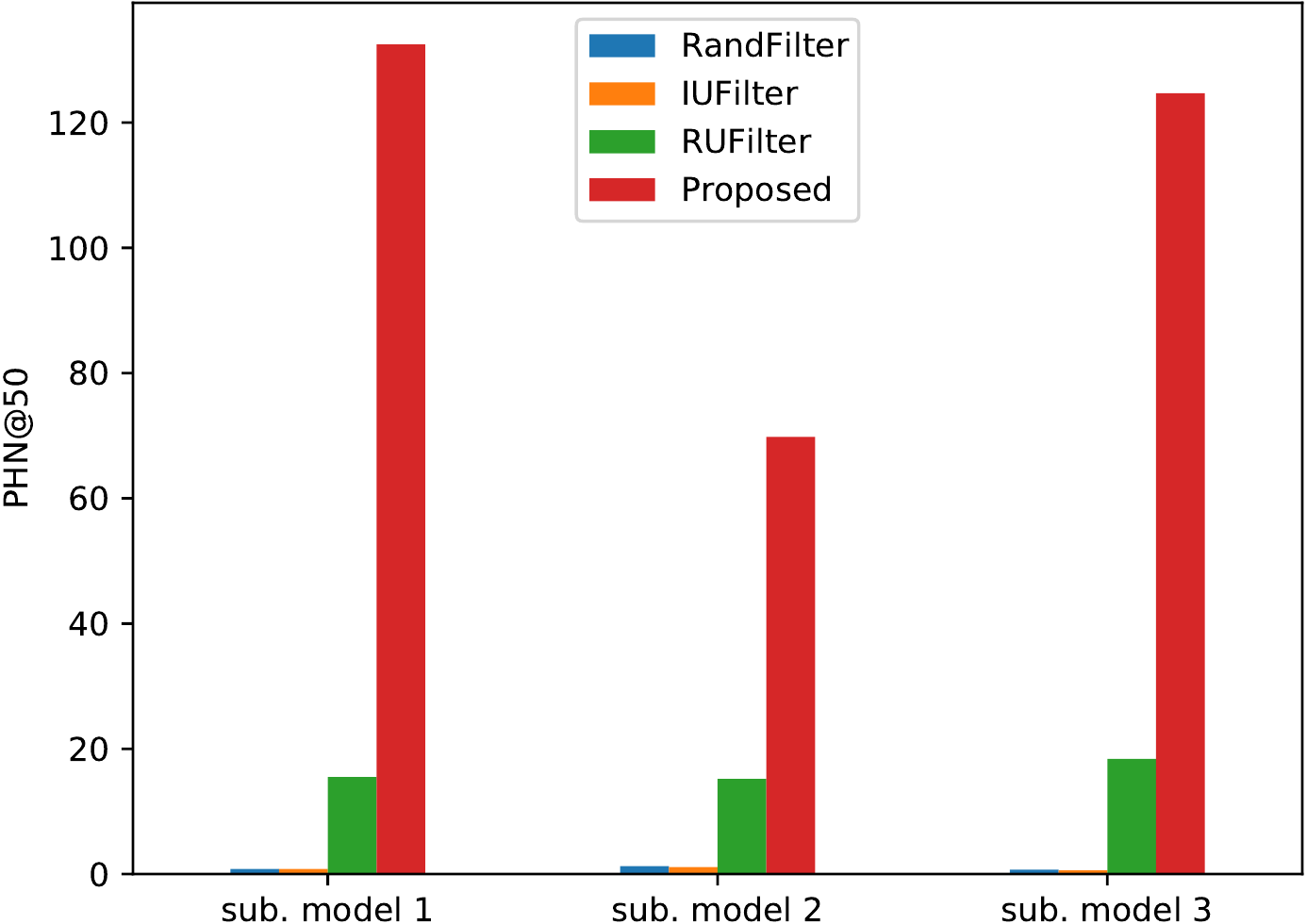}
	\caption{\textbf{Visualization of attack performance comparisons on three different substitute models on Gowalla}.  }
	\label{fig:sub_results_gowalla}
\end{figure}

In Fig.~\ref{fig:sub_results_gowalla}, we compare and visualize the attack performance of different methods on three substitute models (i.e., sub. models 1--3) on Gowalla. Although three substitute models are different from the source model, clearly, we can conclude that the proposed method still achieves satisfactory attack performances.

\subsection{Effectiveness of Node Masking Mechanism}
\begin{figure}[!ht]
	\centering
	\includegraphics[width=8cm, height=4.5cm]{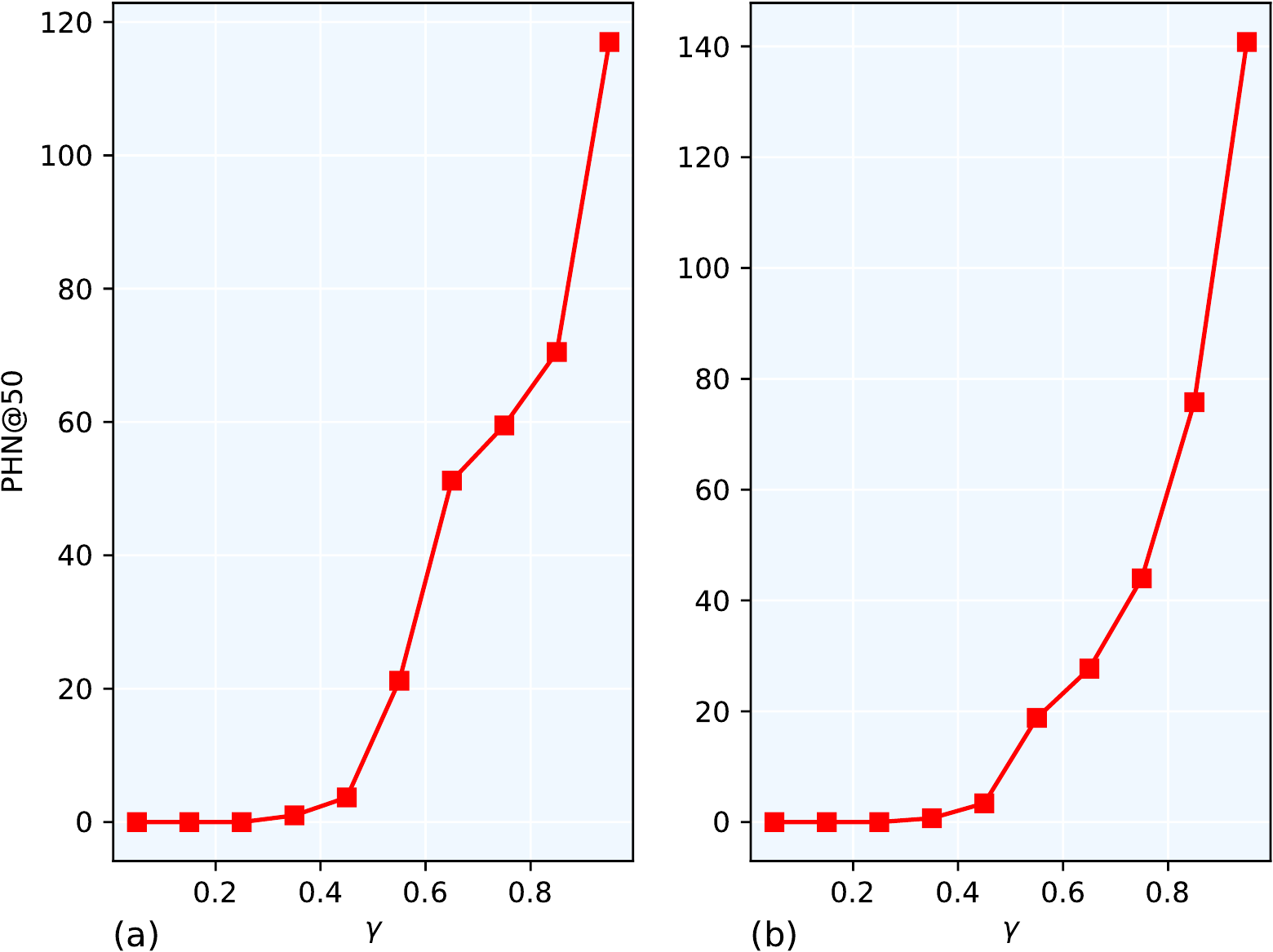}
	\caption{\textbf{Performance variations versus masking thresholds $\gamma$. } (a) and (b) show $PHN@50$ results vs $\gamma$ for target items from $Q_{10}$ on Gowalla and Yelp2018 datasets, respectively.  }
	\label{fig:gamma}
\end{figure}

This section evaluates the performance of our method with different choices of parameter $\gamma$. Specifically, we vary $\gamma$ from 0.05 to 0.95 and compare $PHN@50$. Figure~\ref{fig:gamma} displays the performance curve using target items from $Q_{10}$ item set. 

In Figure~\ref{fig:gamma}, we can see that the item promotion ability of our method is low when we use all users to establish the attack objective (i.e., $\gamma$ as 0). As we increase the masking threshold $\gamma$, $PHN@50$ increases remarkably. This observation confirms the effectiveness of the node masking mechanism.

\subsection{Promoting Items in Retraining Scenes}
In real deployment scenarios, a collaborative filtering model may be retrained from scratch to capture the dynamics of an input graph. We simulate such an item promotion scene by first perturbing a user-item graph followed by retraining a new model on the perturbed graph. We keep all experimental setting the same as that used for the source model.

Table~\ref{tab:promotion_attacks_retrain} reveals that, compared with baseline methods, our method consistently achieves the highest $PHN@50$ with a same perturbation budget. On both datasets, we have about $\boldsymbol{1.4 \times}$ to $\boldsymbol{4.6 \times}$ larger attack ability than the second best method. Therefore, the proposed attack method successfully maintains a strong promotion ability even when we retrain the collaborative filtering model after topological perturbations. Please refer to the Appendix for more results.

\begin{table}[]
\centering
\begin{adjustbox}{width=8.4cm}
\begin{tabular}{|c|l|ccc|}
\hline
\multirow{3}{*}{Dataset}  & \multicolumn{1}{c|}{\multirow{3}{*}{Attack}} & \multicolumn{3}{c|}{Low popularity items}                                                     \\ \cline{3-5} 
                          & \multicolumn{1}{c|}{}                        & \multicolumn{1}{c|}{$Q_{10}$}   & \multicolumn{1}{c|}{$Q_{30}$}   & $Q_{50}$   \\ \cline{3-5} 
                          & \multicolumn{1}{c|}{}                        & \multicolumn{1}{c|}{\textit{PHN@50 ($\Delta_{10}^1$)}} & \multicolumn{1}{c|}{\textit{PHN@50 ($\Delta_{30}^1$)}} & \textit{PHN@50 ($\Delta_{50}^1$)} \\ \hline \hline
\multirow{4}{*}{Gowalla}  & RandFilter                                   & \multicolumn{1}{c|}{8.8}             & \multicolumn{1}{c|}{12.4}            & 14.9            \\ \cline{2-5} 
                          & IUFilter                                     & \multicolumn{1}{c|}{1.9}             & \multicolumn{1}{c|}{5.0}             & 7.8             \\ \cline{2-5} 
                          & RUFilter                                     & \multicolumn{1}{c|}{5.2}             & \multicolumn{1}{c|}{11.3}            & 13.6            \\ \cline{2-5} 
                          & \textbf{Proposed}                            & \multicolumn{1}{c|}{\textbf{23.4}}   & \multicolumn{1}{c|}{\textbf{23.3}}   & \textbf{20.5}   \\ \hline \hline
\multirow{4}{*}{Yelp2018} & RandFilter                                   & \multicolumn{1}{c|}{5.0}             & \multicolumn{1}{c|}{6.4}             & 8.3            \\ \cline{2-5} 
                          & IUFilter                                     & \multicolumn{1}{c|}{1.2}             & \multicolumn{1}{c|}{1.7}             & 3.9             \\ \cline{2-5} 
                          & RUFilter                                     & \multicolumn{1}{c|}{5.1}             & \multicolumn{1}{c|}{3.6}             & 6.4             \\ \cline{2-5} 
                          & \textbf{Proposed}                            & \multicolumn{1}{c|}{\textbf{23.1}}   & \multicolumn{1}{c|}{\textbf{16.7}}   & \textbf{13.0}   \\ \hline
\end{tabular}
\end{adjustbox}
\caption{Performance comparisons of different attacks in improving a target item's popularity on Gowalla and Yelp2018 datasets \textbf{with retraining}. Three low popularity item sets ($Q_{10}, Q_{30}, Q_{50}$) are used for performance evaluation with perturbation budgets as $\Delta_s^1$ $(s=10, 30, 50)$. $PHN@50$ is averaged over 30 randomly selected target items at each item set. The best performances are marked in bold. }
\label{tab:promotion_attacks_retrain}
\end{table}

\section{Conclusion}
\label{conclusion}
In this work, we have proposed a novel view on promoting items in GNN-based collaborative filtering models based on topological attacks. Our formulation identifies users that play pivotal roles to promote a target item. We then propose a node masking mechanism to effectively improve vanilla gradient-based solutions. A resource-efficient approach is developed to make our method scalable to large-scale graphs. Experimental results demonstrate that the proposed method can significantly promote a target item. Our findings raise the security and adversarial robustness issues of GNN-based collaborative filtering models in practical scenes. 

\section*{Appendices}

\subsection*{A. Detailed Training Setups \& Performances of The Main Model and Three Victim Models}
For the main model evaluated in Section 4.2, we utilized the default parameter settings as in the official PyTorch implementation of work \cite{he2020lightgcn}. We also craft three victim models by setting different training hyparameters.

To be more specific, four hyperparameters are changed to obtain three different victim models, i.e., sub. model 1, sub. model 2 and sub. model 3. The hyperparameters are respectively as 1) the number of layers $L$; 2) the length of a user or item embedding $d$; 3) the number of training epochs (i.e., \#epochs); and 4) the random seed used for model randomization. The hyperparameters of the main model and three other victim models are listed in Table ~\ref{tab:setup_victim_models}.  

\begin{table}[!htbp]
\centering
\begin{tabular}{|l|c|c|c|c|}
\hline
             & $L$ & $d$ & \#epochs & seed \\ \hline
main model   & 3   & 64  & 1000     & 2020 \\ \hline
sub. model 1 & 2   & 64  & 200      & 2022 \\ \hline
sub. model 2 & 2   & 128 & 200      & 2022 \\ \hline
sub. model 3 & 3   & 128 & 200      & 2022 \\ \hline
\end{tabular}
\caption{The training setups of the main victim model (for white-box attacks) and three different victim models (for black-box attacks). }
\label{tab:setup_victim_models}
\end{table}

The main victim model and three victim models adopt a same $f_{comb}$ (defined in Eq.~(\ref{eq:combine_embedding})) to combine user or item embeddings as for the main evaluation model, following existing studies \cite{he2020lightgcn, wu2021self}. Namely, the final embedding is formed by a weighted sum of embeddings output from each layer,
\begin{equation}
    \mathbf{z}_u  = \sum_{l=0}^L \alpha_l \; e_u^{(l)},  \quad 
    \mathbf{z}_i  = \sum_{l=0}^L \alpha_l \; e_i^{(l)}
    \label{eq:combine_embedding_final}
\end{equation}
where $\alpha_l$ is set uniformly as $1/(L+1)$.

After sufficient training of the main model and three victim models, we report in Table ~\ref{tab:clean_performance} their recommendation performances on clean Gowalla and Yelp2018 datasets. Similar as the main model, all three victim models can achieve satisfactory recommendation performances, indicating that they may also be deployed in real recommendation scenes. Therefore, we assume these three sub. models as victim models, and we evaluate the attacking performance of our method on them in a black-box attack setting.

\begin{table}[]
\centering
\begin{adjustbox}{width=8.4cm}
\begin{tabular}{|c|l|c|c|c|}
\hline
                          & Model        & \textit{Precision} & \textit{Recall} & \textit{NDCG}  \\ \hline \hline 
\multirow{4}{*}{Gowalla}  & main model   & 0.056     & 0.182  & 0.154 \\ \cline{2-5} 
                          & sub. model 1 & 0.053     & 0.173  & 0.148 \\ \cline{2-5} 
                          & sub. model 2 & 0.055     & 0.180  & 0.153 \\ \cline{2-5} 
                          & sub. model 3 & 0.055     & 0.181  & 0.153 \\ \hline \hline 
\multirow{4}{*}{Yelp2018} & main model   & 0.028     & 0.063  & 0.052 \\ \cline{2-5} 
                          & sub. model 1 & 0.026     & 0.582  & 0.048 \\ \cline{2-5} 
                          & sub. model 2 & 0.028     & 0.062  & 0.051 \\ \cline{2-5} 
                          & sub. model 3 & 0.029     & 0.064  & 0.053 \\ \hline
\end{tabular}
\end{adjustbox}
\caption{The recommendation performances of the pretrained main model and three victim models on \textbf{clean} Gowalla and Yelp2018 datasets.}
\label{tab:clean_performance}
\end{table}

\subsection*{B. Perturbed Graphs Impose Minimal Influences on Overall Recommendation Performances }
In this section, we intends to evaluate whether perturbations on the user/item graphs can influence the overall recommendation performances of a GNN-based collaborative filtering model. On Gowalla and Yelp2018 datasets, the evaluated graphs are generated on $Q_{10}$ items with the number of perturbation budgets set as 12 and 25, respectively. 

The recommendation performances of four models are reported in Table ~\ref{tab:perturbed_performance}. Compared with those in Table ~\ref{tab:clean_performance}, the recommendation performances almost stay unchanged by using perturbed graphs generated by our method. This is because we are utilizing a very small perturbation budget compared to the (large) number of users in the recommendation system, thus leading to minimal influence to the overall recommendation performances. For example, on Gowalla, we perturb 12 users which only takes about 0.04\% of all users. The minimal influences on overall recommendation performances imply that the model will perform almost the same to recommend all other items, which also indicates difficulties to detect the existence of adversarial perturbations by our method.

\begin{table}[]
\centering
\begin{adjustbox}{width=8.4cm}
\begin{tabular}{|c|l|c|c|c|}
\hline
                          & Model        & \textit{Precision} & \textit{Recall} & \textit{NDCG}  \\ \hline \hline 
\multirow{4}{*}{Gowalla}  & main model   & 0.056     & 0.182  & 0.154 \\ \cline{2-5} 
                          & sub. model 1 & 0.052     & 0.173  & 0.148 \\ \cline{2-5} 
                          & sub. model 2 & 0.055     & 0.180  & 0.153 \\ \cline{2-5} 
                          & sub. model 3 & 0.055     & 0.181  & 0.153 \\ \hline \hline 
\multirow{4}{*}{Yelp2018} & main model   & 0.028     & 0.063  & 0.052 \\ \cline{2-5} 
                          & sub. model 1 & 0.026     & 0.580  & 0.048 \\ \cline{2-5} 
                          & sub. model 2 & 0.028     & 0.062  & 0.051 \\ \cline{2-5} 
                          & sub. model 3 & 0.029     & 0.064  & 0.053 \\ \hline
\end{tabular}
\end{adjustbox}
\caption{The recommendation performances of the pretrained main model and three victim models on \textbf{perturbed} graphs on Gowalla and Yelp2018 datasets. The perturbed graphs are generated using the proposed method with low popularity items $Q_{10}$ and $\Delta$ as 12 and 25, respectively on Gowalla and Yelp2018 datasets.}
\label{tab:perturbed_performance}
\end{table}

\subsection*{C. More Results for Retraining Scenes}
In Table~\ref{tab:promotion_attacks_retrain_Q}, we present the item promotion results on Yelp2018 of low popularity items, i.e., $Q_{10}, Q_{30}, Q_{50}$ with perturbations as $\Delta_{10}^1, \Delta_{30}^1, \Delta_{50}^1$ and $\Delta_{10}^2, \Delta_{30}^2, \Delta_{50}^2$ by retraining models on perturbed user and item graphs. Consistent with results reported for Gowalla in Table~\ref{tab:promotion_attacks_retrain}, the proposed method outperforms baseline methods by a large margin in terms of the $PHN@50$ metric.  

In addition, we evaluate and compare the $PHN@50$ values of the retrained GNN-based collaborative filtering model for two sets of items: the first set of items are clean items (i.e., no perturbations), while the second set are target items with certain perturbations. For each set of items, we randomly selected 30 items and report their averaged $PHN@50$ values.

\begin{figure}
     \centering
     \begin{subfigure}[b]{0.42\textwidth}
         \centering
         \includegraphics[width=\textwidth]{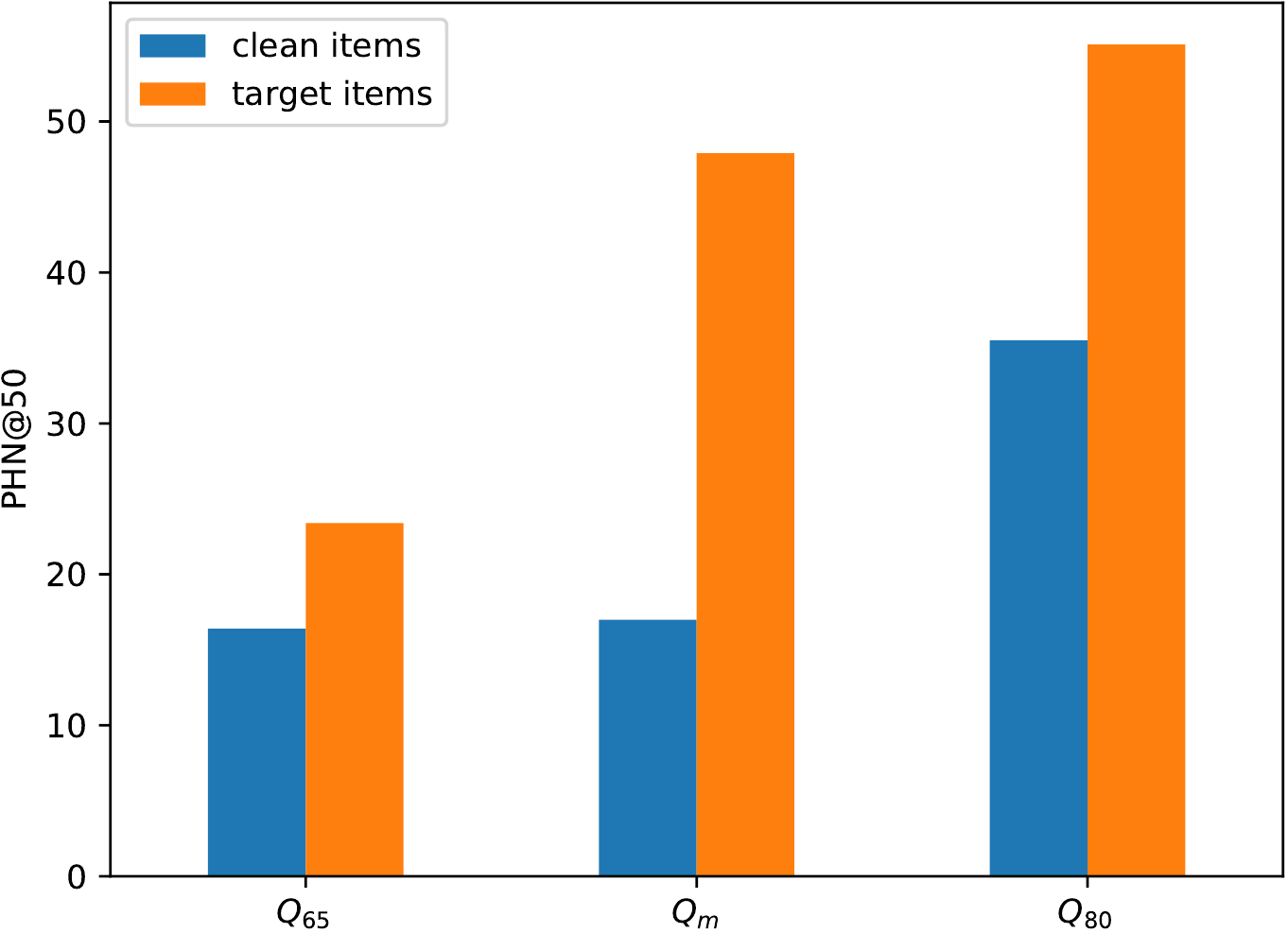}
         \caption{$PHN@50$ comparisons on Gowalla}
         \label{fig:clean_vs_retrain_gowalla}
     \end{subfigure}
     \hfill
     \begin{subfigure}[b]{0.42\textwidth}
         \centering
         \includegraphics[width=\textwidth]{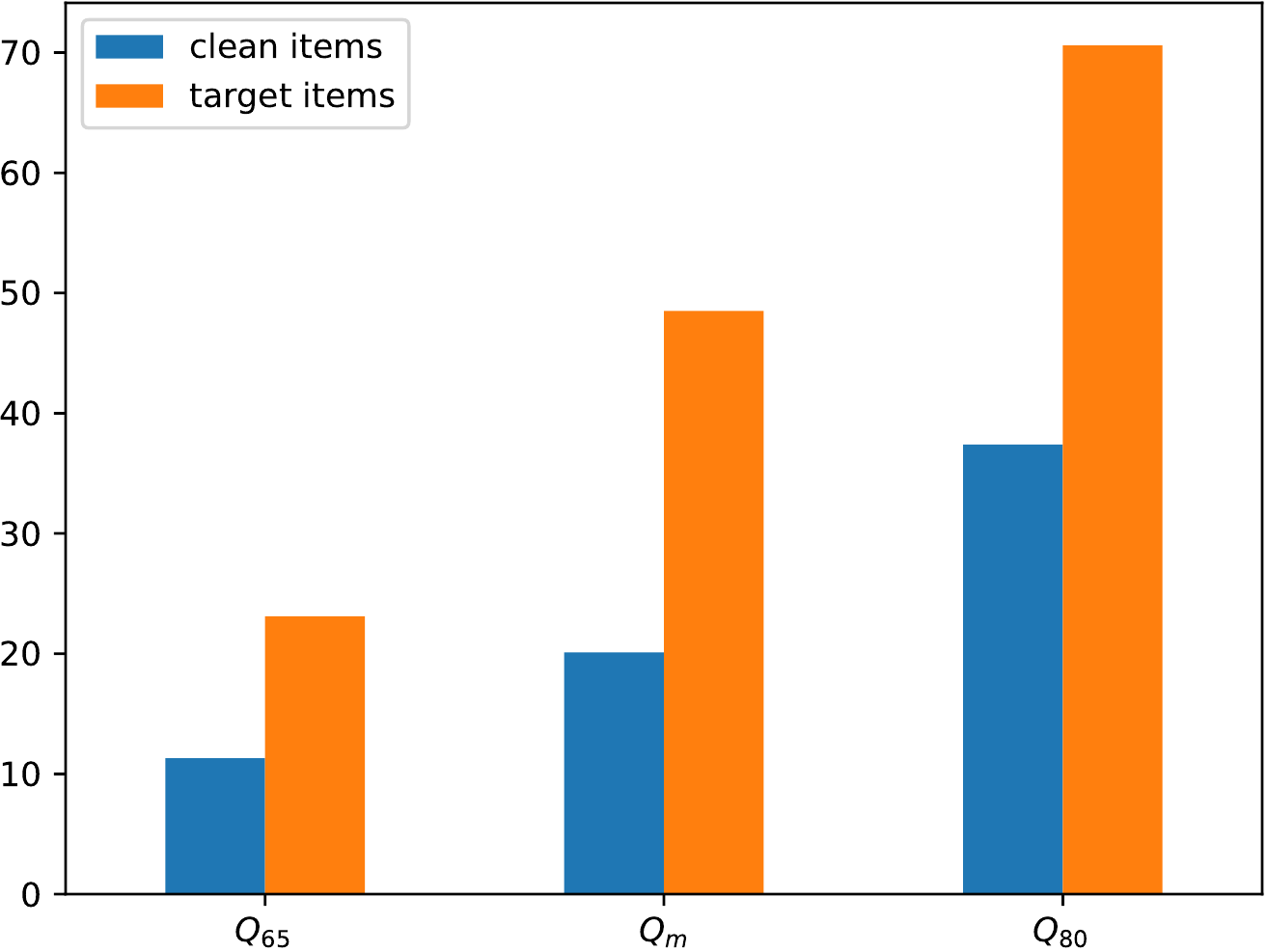}
         \caption{$PHN@50$ comparisons on Yelp2018}
         \label{fig:clean_vs_retrain_yelp}
     \end{subfigure}
    \caption{Performance comparisons between items from clean and promoted target item sets on Gowalla and Yelp2018 datasets.}
    \label{fig:clean_vs_retrain}
\end{figure}

In Fig.~\ref{fig:clean_vs_retrain}, we report the comparison results on Gowalla and Yelp2018, where the target items are from the $Q_{10}$ set. The clean items are from $Q_{65}, Q_m$, and $Q_{80}$, respectively. The perturbation budgets are computed by subtracting degrees of clean items from those of $Q_{10}$ items. To be specific, as reported in Table~\ref{tab:perturb_budgets}, the degree differences are $\Delta_{10}^1$ and  $\Delta_{10}^2$ to promote items from $Q_{10}$ set to $Q_{65}, Q_m$ sets; and the perturbation budgets are 14 and 33 to promote items from $Q_{10}$ to $Q_{80}$.

From Fig.~\ref{fig:clean_vs_retrain}, we observe that target items display clearly higher $PHN@50$ values than their clean counterparts even with the same number of degrees. This phenomena reveals that the proposed attack has indeed identified potential users that will have significant impacts on target items regardless whether the collaborative filtering will be trained or not.

\begin{table}[]
\centering
\begin{adjustbox}{width=8.4cm}
\begin{tabular}{|c|l|ccc|}
\hline
\multirow{3}{*}{Dataset}  & \multicolumn{1}{c|}{\multirow{3}{*}{Attack}} & \multicolumn{3}{c|}{Low popularity items}                                                     \\ \cline{3-5} 
                          & \multicolumn{1}{c|}{}                        & \multicolumn{1}{c|}{$Q_{10}$}   & \multicolumn{1}{c|}{$Q_{30}$}   & $Q_{50}$   \\ \cline{3-5} 
                          & \multicolumn{1}{c|}{}                        & \multicolumn{1}{c|}{\textit{PHN@50 ($\Delta_{10}^2$)}} & \multicolumn{1}{c|}{\textit{PHN@50 ($\Delta_{30}^2$)}} & \textit{PHN@50 ($\Delta_{50}^2$)} \\ \hline \hline
\multirow{4}{*}{Gowalla}  & RandFilter                                   & \multicolumn{1}{c|}{11.6}             & \multicolumn{1}{c|}{11.2}            & 14.3            \\ \cline{2-5} 
                          & IUFilter                                     & \multicolumn{1}{c|}{1.7}             & \multicolumn{1}{c|}{3.8}             & 7.8             \\ \cline{2-5} 
                          & RUFilter                                     & \multicolumn{1}{c|}{9.6}             & \multicolumn{1}{c|}{15.9}            & 19.0            \\ \cline{2-5} 
                          & \textbf{Proposed}                            & \multicolumn{1}{c|}{\textbf{47.9}}   & \multicolumn{1}{c|}{\textbf{49.5}}   & \textbf{46.6}   \\ \hline \hline
\multirow{4}{*}{Yelp2018} & RandFilter                                   & \multicolumn{1}{c|}{6.3}             & \multicolumn{1}{c|}{9.9}             & 11.9            \\ \cline{2-5} 
                          & IUFilter                                     & \multicolumn{1}{c|}{1.2}             & \multicolumn{1}{c|}{2.0}             & 3.7             \\ \cline{2-5} 
                          & RUFilter                                     & \multicolumn{1}{c|}{6.6}             & \multicolumn{1}{c|}{6.0}             & 8.3             \\ \cline{2-5} 
                          & \textbf{Proposed}                            & \multicolumn{1}{c|}{\textbf{48.5}}   & \multicolumn{1}{c|}{\textbf{35.2}}   & \textbf{25.8}   \\ \hline
\end{tabular}
\end{adjustbox}
\caption{Performance comparisons of different attacks in improving a target item's popularity on Gowalla and Yelp2018 datasets \textbf{with retraining}. Three low popularity item sets ($Q_{10}, Q_{30}, Q_{50}$) are used for performance evaluation with perturbation budgets as $\Delta_s^2 (s=10, 30, 50)$. $PHN@50$ is averaged over 30 randomly selected target items at each item set. The best performances are marked in bold. }
\label{tab:promotion_attacks_retrain_Q}
\end{table}

\subsection*{D. Black-box Attacks on Three Victim Models on Yelp2018 }
\begin{figure}[!ht]
	\centering
	\includegraphics[width=0.45\textwidth]{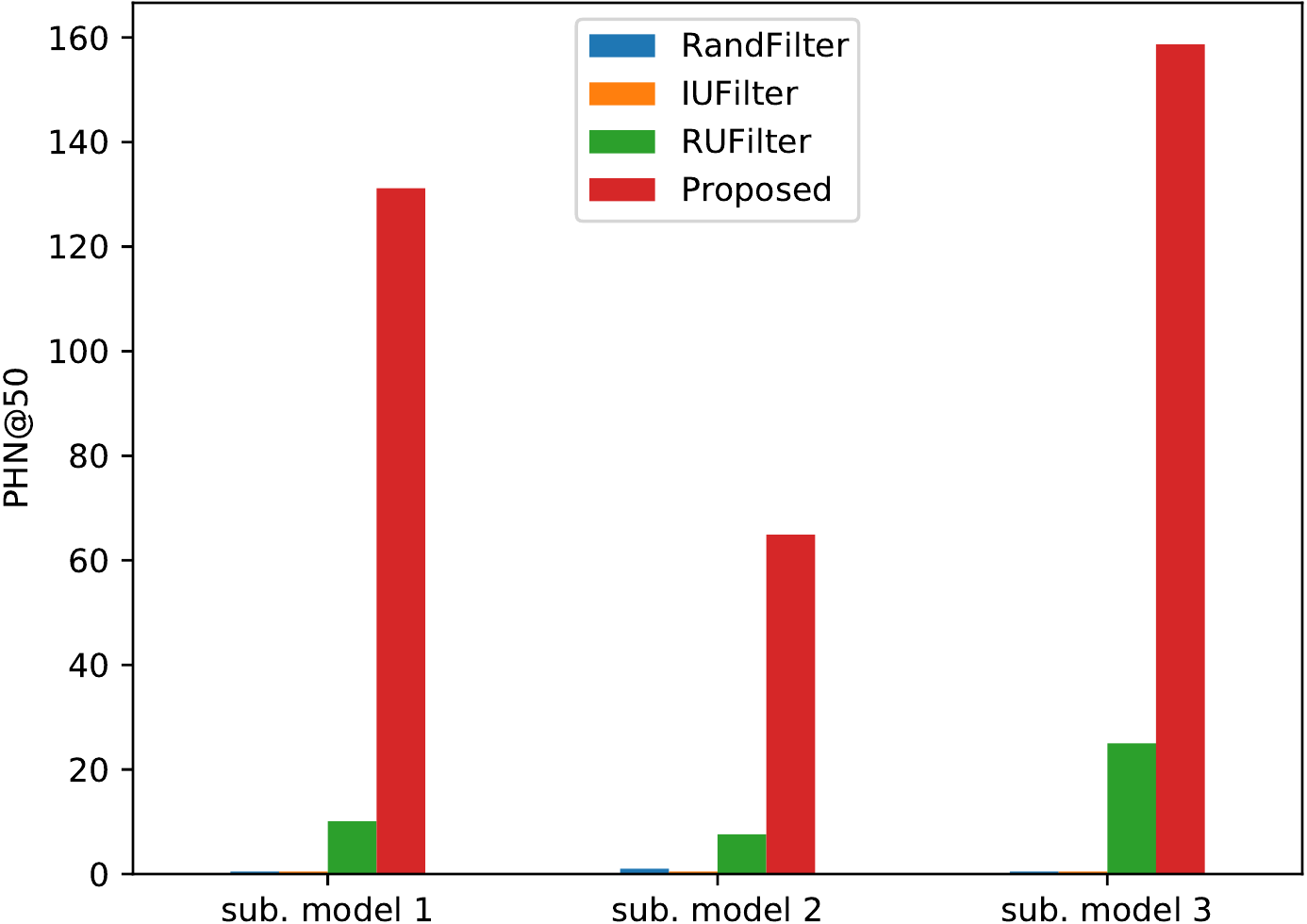}
	\caption{\textbf{Visualization of attack performance comparisons on three different substitute  models on Yelp2018}.  }
	\label{fig:sub_results_yelp}
\end{figure}

In Fig.~\ref{fig:sub_results_yelp}, we show the attack performance comparisons of item promotion methods on three different victim GNN-based collaborative filtering models on Yelp2018 dataset.

For three substitute models, our attack method maintains high item promotion ability, while three baseline methods show little promotion ability on Yelp2018. This observation is consistent with that on Gowalla as visualized in Fig.~\ref{fig:sub_results_gowalla}.

\subsection*{E. Evaluation with Different $PHN@K$ Metrics}
In addition to utilizing the $PHN@K$ with $K$ set as 50 in previous sections, we also report $PHN@10$ and $PHN@20$ in Table~\ref{tab:promotion_attacks_different_K}. Indeed, we can see that the item promotion ability of our method can also outperform baseline methods substantially on $PHN@10$ and $PHN@20$ metrics.

\begin{table}[!htbp]
\centering
\begin{adjustbox}{width=6.8cm}
\begin{tabular}{|c|l|c|c|}
\hline
Dataset                   & Attack            & \multicolumn{1}{l|}{\textit{PHN@10 ($\Delta_{10}^1$)}} & \multicolumn{1}{l|}{\textit{PHN@20 ($\Delta_{10}^1$)}} \\ \hline \hline
\multirow{4}{*}{Gowalla}  & RandFilter        & 0                             & 0                             \\ \cline{2-4} 
                          & IUFilter          & 0                             & 0                             \\ \cline{2-4} 
                          & RUFilter          & 3.9                           & 6.7                           \\ \cline{2-4} 
                          & \textbf{Proposed} & \textbf{18.3}                 & \textbf{41.3}                 \\ \hline \hline
\multirow{4}{*}{Yelp2018} & RandFilter        & 0                             & 0                             \\ \cline{2-4} 
                          & IUFilter          & 0                             & 0                             \\ \cline{2-4} 
                          & RUFilter          & 4                             & 6.5                           \\ \cline{2-4} 
                          & \textbf{Proposed} & \textbf{43.4}                 & \textbf{70.6}                 \\ \hline
\end{tabular}
\end{adjustbox}
\caption{Evaluating performances using different $PHN@K$ metrics. The best performances are marked in bold.}
\label{tab:promotion_attacks_different_K}
\end{table}

% \balance
\bibliographystyle{ACM-Reference-Format}
\bibliography{sample}

% \appendix
% \clearpage

\end{document}